\documentclass[preprint2]{aastex631}

\usepackage{amsmath}
\usepackage{tabularx}
\usepackage{mathrsfs}
\usepackage{xcolor}
\usepackage{CJK}
\usepackage[caption=false]{subfig}

\shorttitle{Rogue Planet}
\shortauthors{Huang et al.}

\graphicspath{{./}{figures/}}

\begin{document}

\begin{CJK*}{UTF8}{gbsn}
%\title{A Rogue Planet Populated the Distant Kuiper Belt}
\title{A Rogue Planet Helps Populate the Distant Kuiper Belt}

%\correspondingauthor{August Muench}
%\email{greg.schwarz@aas.org, gus.muench@aas.org}

\author[0000-0003-1215-4130]{Yukun Huang (黄宇坤)}
\author[0000-0002-0283-2260]{Brett Gladman}
\author[0000-0001-6597-295X]{Matthew Beaudoin}
\affiliation{Department of Physics and Astronomy,
University of British Columbia \\
6224 Agricultural Road,
Vancouver, BC V6T 1Z1, Canada}
\author{Kevin Zhang}
\affiliation{Department of Physics, Cornell University, 109 Clark Hall, Ithaca NY, 14853, USA}

\begin{abstract}
The orbital distribution of transneptunian objects (TNOs) in the distant Kuiper Belt (with semimajor axes beyond the 2:1 resonance, roughly $a$~=~50--100 au) provides constraints on the dynamical history of the outer solar system. Recent studies show two striking features of this region: 1) a very large population of objects in distant mean-motion resonances with Neptune, and 2) the existence of a substantial detached population (non-resonant objects largely decoupled from Neptune). Neptune migration models are able to implant some resonant and detached objects during the planet migration era, but many fail to match a variety of aspects of the orbital distribution. In this work, we report simulations carried out using an improved version of the GPU-based code GLISSE, following 100,000 test particles per simulation in parallel while handling their planetary close encounters. We demonstrate for the first time that a 2 Earth--mass rogue planet temporarily present during planet formation can abundantly populate both the distant resonances and the detached populations, surprisingly even without planetary migration. We show how weak encounters with the rogue greatly increase the efficiency of filling the resonances, while also dislodging TNOs out of resonance once they reach high perihelia. The rogue's secular gravitational influence simultaneously generates numerous detached objects observed at all semimajor axes. These results suggest that the early presence of additional planet(s) reproduces the observed TNO orbital structure in the distant Kuiper Belt.

\end{abstract}


\section{Introduction} \label{sec:intro}
\end{CJK*}

The heavily studied main Kuiper Belt has semimajor axes smaller than the 2:1 resonance at 48 au (often taken to be the outer boundary of the classical belt). 
Beyond the 2:1, the transneptunian region seems not as abundantly populated
and is dominated by  large-eccentricity ($e$)  transneptunian objects (TNOs) 
in the scattering
\citep{Trujillo.2000, Lawler.2018ic}, 
resonant \citep{Gladman.2012, Crompvoets.2022}, and
detached \citep{Gladman.2008}
populations.
This apparent drop in TNO number is partly due to the observational bias that penalizes orbits with large-$a$, large-perihelia ($q$), and large-inclinations ($i$). 
Deriving the intrinsic TNO orbital distribution at large semimajor axis requires well-characterized surveys that properly handle observation bias. 
Modern surveys like CFEPS \citep{Petit.2011}, OSSOS \citep{Bannister.2018}, and 
DES \citep[Dark Energy Survey;][]{Bernardinelli.2022}
all show evidence for an abundant population of $a$~=~50--100 au TNOs (referred to as `the distant belt' here).
Studies that accounted for this bias \citep{Gladman.2012, Pike.2015, Volk.2018} all concluded that the distant resonances are heavily populated.
The distant $n$:1 resonances are particularly crowded, with populations comparable to the closer 3:2 \citep{Crompvoets.2022}. 
Similar estimates indicate that the detached region hosts at least as many TNOs as the hot classical belt 
\citep[][Beaudoin et al. 2022, submitted to PSJ]{Petit.2011}. 
All evidence points to an abundantly populated distant Kuiper Belt whose inventory should be greatly improved by LSST \citep{Collaboration.2009}.

Neptune migration models have been proposed to create the distant resonant and detached populations. 
\citet{Hahn.2005} simulated Neptune's smooth outward migration into both dynamically cold and heated disks;
neither case populates the distant resonances as much as the 3:2 and 2:1. 
\citet{Gomes.2008} realized detached objects can be created during Neptune's migration via Kozai $q$ lifting. 
Grainy Neptune migrations, in which Neptune's $a$ jumps due to planet encounters \citep{Nesvorny.2016, Kaib.2016l9v}, are also able to capture some scattering particles into the distant resonances.  
\citet{Pike.2017} bias a Nice model simulation from \citet{Brasser.2013} (where Neptune undergoes a high-$e$ phase during outward migration) using the OSSOS survey simulator
and conclude this model doesn't produce large-enough populations for many 
distant resonances. 
\citet{Crompvoets.2022} suggest their recent  resonant-population estimates disfavor all migration models, as they underpopulate the $n$:1 and $n$:2 resonances; instead an underlying sticking of scattering TNOs to the resonances is preferred, although the efficiency is too low \citep{Yu.2018}.

Perhaps effects other than migration are important. Passing stars, even in very dense initial stellar birth cluster environment, are ineffective perturbers inside 200~au \citep{Brasser.2015, Batygin.2020jr}. One way to create high-$q$ TNOs is via the presence of additional mass(es), whose secular gravitational effect elevates objects from the scattering into the detached 
population. The initial creation and scattering of now-gone planetary-scale objects in the outer solar system is reasonable \citep[\textit{egs.}][]{Stern.1991, Chiang.2005, Silsbee.2018, Gladman.2021}.
\citet{Gladman.2002} postulated that additional planetary-mass bodies could account for large-$q$ detached objects like 2000 CR$_{105}$.
This evolved into the `Rogue Planet' hypothesis \citep{Gladman.2006} in which an Earth-scale Neptune-crossing rogue planet (initially starting on a low eccentricity
orbit) temporarily present in the early solar system creates detached TNOs, even as far out as Sedna; they showed that the perihelion-lifting effect is dominated by the single most-massive object, which shares the typical 100-Myr dynamical lifetime of Neptune-scattering bodies.
\citet{Lykawka.2008} also proposed a resident trans-Plutonian planet (with $a=$~100--175 au, $q>80$ au, and 0.3--0.7 $M_\oplus$) to sculpt the Kuiper Belt and generate a substantial population of detached TNOs.

In the context of solar system studies, `rogue planets' refers to planets born in our solar system that are scattered away from their formation location and could have left behind orbital structures  caused by their temporary presence. We point out the same terminology is sometimes also applied to interstellar free-floating planets. 

In light of the additional high-$q$ TNO discoveries in the last 15 years, we
revisit the rogue planet hypothesis.
We show that a rogue planet temporarily present on an eccentric orbit sufficiently populates both the resonant and detached populations in the $a$~=~50--100 au Kuiper Belt, even without any planetary migration. 

\vspace*{1mm}

\section{Dynamics from the 4 Giant Planets}\label{sub:ref}

\begin{figure*}[htb!]
  \centering
  \includegraphics[width=1\textwidth]{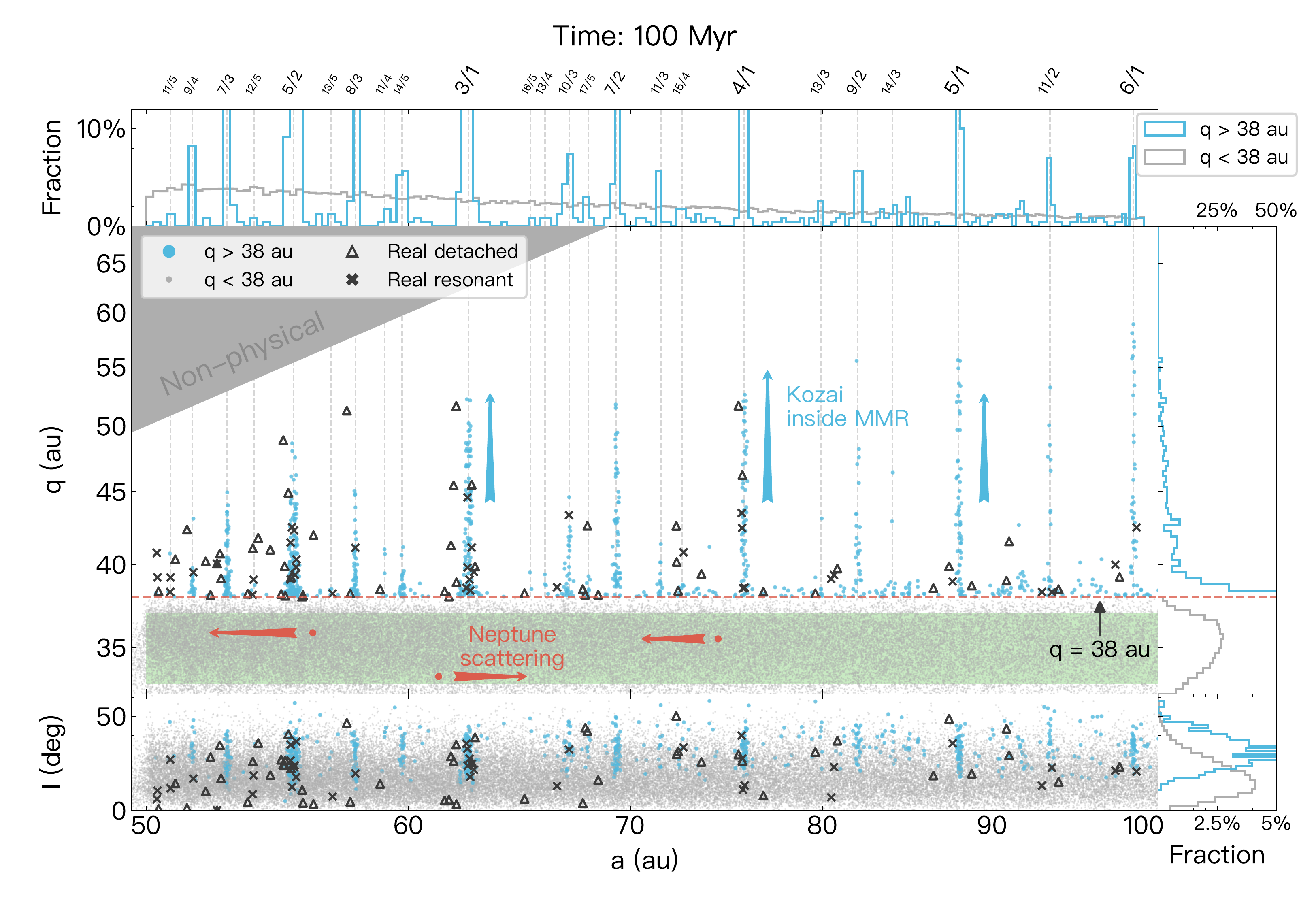}
  \caption{$a, q, i$ distributions of TNOs at 100 Myr in the reference simulation (animated in the online version). 
  The initial conditions have $33 < q_0 < 37$ au (shaded green region). 
  Arrows in the $a,q$ panel illustrates the two dominant dynamical effects in the scattering disk: horizontal Neptune scattering (red arrows) and vertical resonant $q$ lifting (blue arrows). 
  The three side panels show histograms for simulated $q < 38$ au (gray) and  $q > 38$ au (blue) particles. 
  At the simulation's end, only $\simeq$1\% of $a$~=~ 50--100~au particles have $q > 38$ au, with the vast majority being resonant objects and almost none being detached objects.
  The superimposed real $q>38$~au detached objects (black triangles) show an apparent concentration near resonances (especially $n$:1 and $n$:2, labelled at top). 
  }
  \label{fig:ref}
\end{figure*}

To quantify the dynamical effects in the distant Kuiper Belt induced by the four giant planets alone, we show a reference simulation with a 
synthetic young scattering disk. 
100,000 test particles, starting from $a$=50~au,  were placed following a $dN/da \propto a^{-2.5}$ 
distribution.\footnote{We showed this with a simple Neptune scattering simulation in which initially low-$e$ and low-$i$ objects near Neptune followed an $a^{-2.5}$ distribution 
at $\sim$50~Myr.
This is steeper than the longer-term steady-state of 
$dN/da \propto a^{-1.5}$, predicted by a diffusion
approximation 
%in which TNOs random walk in semimajor axes with constant perihelia 
\citep{Yabushita.1980} and validated by cometary dynamics simulations \citep{Levison.1997}.
}
A uniform $q_0$=33--37~au distribution was used to cover the current values of scattering objects, but will allow us to post-facto explore the early scattering disk's parameters integration by weighting the $q_0$ values
(Sec.~\ref{sub:biases}).
The initial inclination distribution follows $\sin{i}$ times a gaussian of $15^\circ$ width, the same distribution as the hot main Kuiper Belt objects \citep{Brown.2001, Petit.2011}. 
All phase angles ($\Omega$, $\omega$, and $M$) are random and orbital
elements are always converted to the J2000 barycentric frame.

We integrated for 100 Myr, with 4 giant planets on their current orbits,  using a regularized version of \textsc{Glisse} \citep{Zhang.2022}. 
This modified integrator \textsc{Glisser} can propagate $\sim$10$^5$ test particles on a GPU, while resolving close encounters with planets on multiple CPU cores using many \textsc{Swift} subroutine calls \citep{Levison.1994}. We have verified this integrator in several common test problems, confirming it correctly handles the resonant dynamics, secular dynamics and scattering dynamics. 
\textsc{Glisser} provides final orbital distributions statistically identical to those simulated by other standard orbital integrators like \textsc{Mercury} \citep{Chambers.1999} and \textsc{Swift} \citep{Levison.1994}.

The 100~Myr snapshot for the reference simulation's movie is shown in Fig.~\ref{fig:ref}. 
We limit our comparisons to $a$~=~50--100~au because this region has a meaningful density of known $q>$~38~au resonant and detached objects, 
making a comparison feasible. 
With four giant planets, two dynamical processes dominate the scattering disk. 
At $q<$~38~au, TNOs are steadily scattered due to 
their proximity at perihelion passages to Neptune's orbit; 
this produces horizontal movement (denoted by red particles and arrows in Fig.~\ref{fig:ref} for a few examples) on the $(a,q)$ plot as scattering TNOs random walk in $a$ while approximately preserving $q$. 
At larger perihelia, where weaker Neptunian encounters less effectively change the TNO's orbital elements, the dominant dynamics occurs at Neptunian mean-motion resonances. 
The resonances allow evolution to higher-$q$ and higher-$i$ orbits via the Kozai-Lidov mechanism inside mean-motion resonances \citep{KOZAI.1962, Gomes.2008}. 
The perihelion evolution (blue dots and arrows in Fig.~\ref{fig:ref}) is clearly stronger 
along $n$:1 and $n$:2 resonances. 
Unfortunately, the overall efficiency of the resonant $q$-lifting effect is low, with only $\approx$$1\%$ of $a$~=~50--100~au objects reaching $q > 38$~au in 100 Myr. 
Furthermore, we determined that almost every particle located in Fig.~\ref{fig:ref}'s resonant spikes 
was {\it initially} within $\pm$0.3~au of the corresponding resonant center meaning they by chance started resonant rather than being delivered to it. 
This indicates that resonant sticking \citep[a mechanism characterized by scattering objects evolving through intermittent temporary resonance captures,][]{Lykawka.2007} is not the main source of the high-$q$ resonant TNOs. 
We will return to this in Sec.~\ref{sub:rogue}.

Compared to real TNOs in the same region (black triangles and crosses in Fig.~\ref{fig:ref}), this reference model produces some resonant objects but 
barely any detached objects, especially between the resonances at high $q$. 
This mismatch is unsurprising because Neptune with a largely unchanging orbit is extremely inefficient at detaching objects from the scattering disk \citep{Gladman.2002}. 
Although resonance escape can (rarely) happen at high $q$ without Neptune migration, \citet[fig. 10]{Gomes.2008} conclude that Neptune migration is needed to break the  reversibility.
Therefore, grainy migration models \citep[\textit{egs.}][]{Nesvorny.2016, Kaib.2016l9v} introduced moderate ($\sim$0.1~au) semimajor axis jumps to Neptune's migration history in order to detach objects from resonances
by suddenly moving the resonance borders. 
\citet{Nesvorny.2016} show how grainy Neptune migration results in greater resonance
trapping and although they do not bias their numerical results to see if the orbital distribution matches known TNOs, their detached population agrees with the recent
observational measurement (Beaudoin et al. 2022, submitted).

\begin{figure*}[htb!]
  \centering
  \includegraphics[width=1\textwidth]{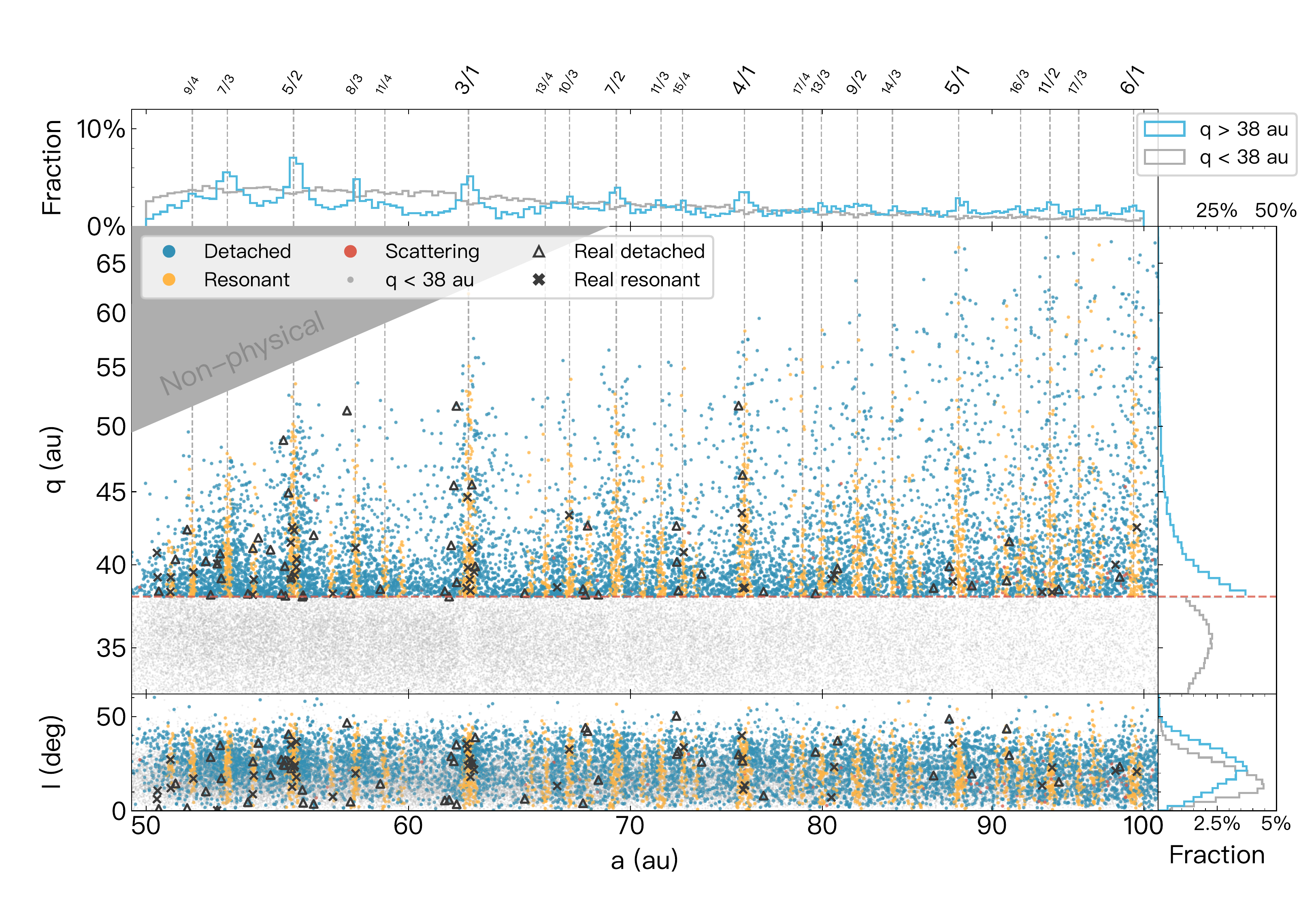}
  \caption{$a, q, i$ distributions and histograms of test particles under 100-Myr gravitational influences of a 2 $M_\oplus$ rogue (initial $a_r=300$ au, $q_r=40$ au, and $i_r = 20^\circ$) and the 4 giant planets. 
  The online animated version shows the entire evolution.
  Particles are color-coded based on their constantly-evolving dynamical classes when they are above the red dashed line of $q = 38$ au. With a rogue present, the resonant population (yellow) is now 3 times larger than in the reference simulation. The detached objects (blue) are created across all semimajor axes, with the high-$q$ ones concentrated near resonances (see the $a$ histogram at the top). Depletions at strong resonances are visible in the $q<38$~au scattering disk (gray), due to the rogue boosting the efficiency of resonant $q$ lifting.}
  \label{fig:main}
\end{figure*}

\section{Dynamical Effects Induced by the Rogue Planet}\label{sub:rogue}

The Letter presents a proof of concept that a temporarily present planet (called a {\it rogue}) can create high-perihelion objects distributed similarly to the observed Kuiper Belt, with sufficient efficiency to match observations and comparable to grainy migration simulations.
We added a $m_r=2 M_\oplus$ rogue with an initial $a_r= 300$ au, $q_r= 40$ au, and $i_r=20^\circ$ orbit to the simulation, and integrated it with the same 100,000 test particles to 100 Myr. 
The chosen rogue parameters (mass, semimajor axis, and dynamical lifetime) were inspired by the preliminary study of \citet{Gladman.2006} where the authors demonstrated such a rogue detaches objects from the scattering disk through secular $q$ forcing,
but they
had insufficient statistics to examine the rogue's role in populating distant resonances and detaching objects from these resonances (which we find is the major dynamical mechanism populating $a$~=~50--100 au). 
We set the rogue's $q_0$=40~au to produce weak $a_r$ mobility over the simulation, as we're concentrating on the new dynamics the rogue brings to the 
distant belt,
rather than exploring the 
enormous parameter space of possible rogue histories.
The TNO orbital evolution  
is displayed in Fig.~\ref{fig:main}. 

Each particle in Fig.~\ref{fig:main}
is categorized into one of three dynamical classifications of detached (blue), resonant (orange), and scattering (red), 
using its 10-Myr dynamical history \citep{Gladman.2008} around a particular 
moment\footnote{For example, the dynamical class at 100 Myr is based on the orbital history from 95--105 Myr. Classification was performed at each 0.14~Myr output interval (except for the first 5 Myr).}
in the animated version of Fig.~\ref{fig:main}. 
We encourage the reader watch the movie on the journal website, which shows the constantly-evolving dynamical classes of each test particles. Only $q > 38$ au particles are color-coded based on these classes; the $q < 38$ au particles (gray) are less relevant to the problem we are exploring, as their distribution is largely set by initial conditions. 

One striking difference in Fig.~\ref{fig:main} is that the rogue's secular effect detaches TNOs directly from the scattering disk across all semimajor axes. 
This lifting is faster at larger $a$;
for TNOs with $a \ll a_r$ and orbital period $P$, the order-of-magnitude $q$ oscillation timescale $P_\text{sec}$ induced by the rogue is given by \citep{Gladman.2006}:
\begin{equation}\label{eq:P_sec}
    \frac{P_\text{sec}}{P} \sim \left(\frac{M_\odot}{m_r}\right) \left(\frac{a_r}{a}\right)^{3} \left(1-e_r^2\right)^{3/2},
\end{equation}
where 
$e_r$ is the rogue's eccentricity. 
For a 2 $M_\oplus$ rogue with $a_r\simeq300$~au and $e_r\simeq0.87$, $P_\text{sec}$ for $a=50$--100 au varies from $\simeq$1.5~Gyr to 500~Myr, longer than the $\sim$100 Myr dynamical lifetime of the rogue. 

\begin{figure*}[htb!]
\centering
    \subfloat[No rogue]{
      \centering
      \includegraphics[width=0.5\textwidth]{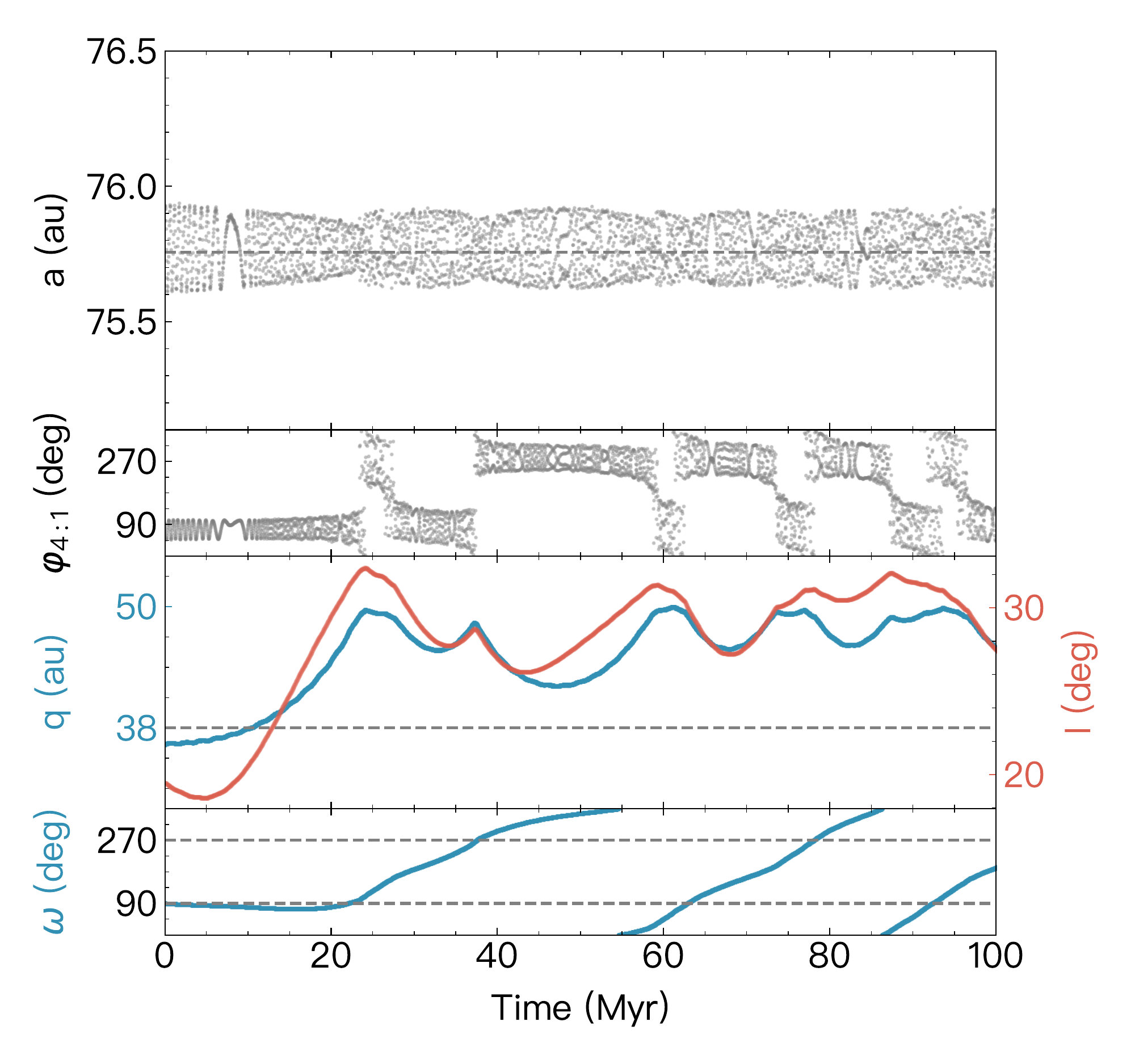}
      \label{subfig:par_norogue}
    }
    \subfloat[With rogue]{
      \centering
      \includegraphics[width=0.5\textwidth]{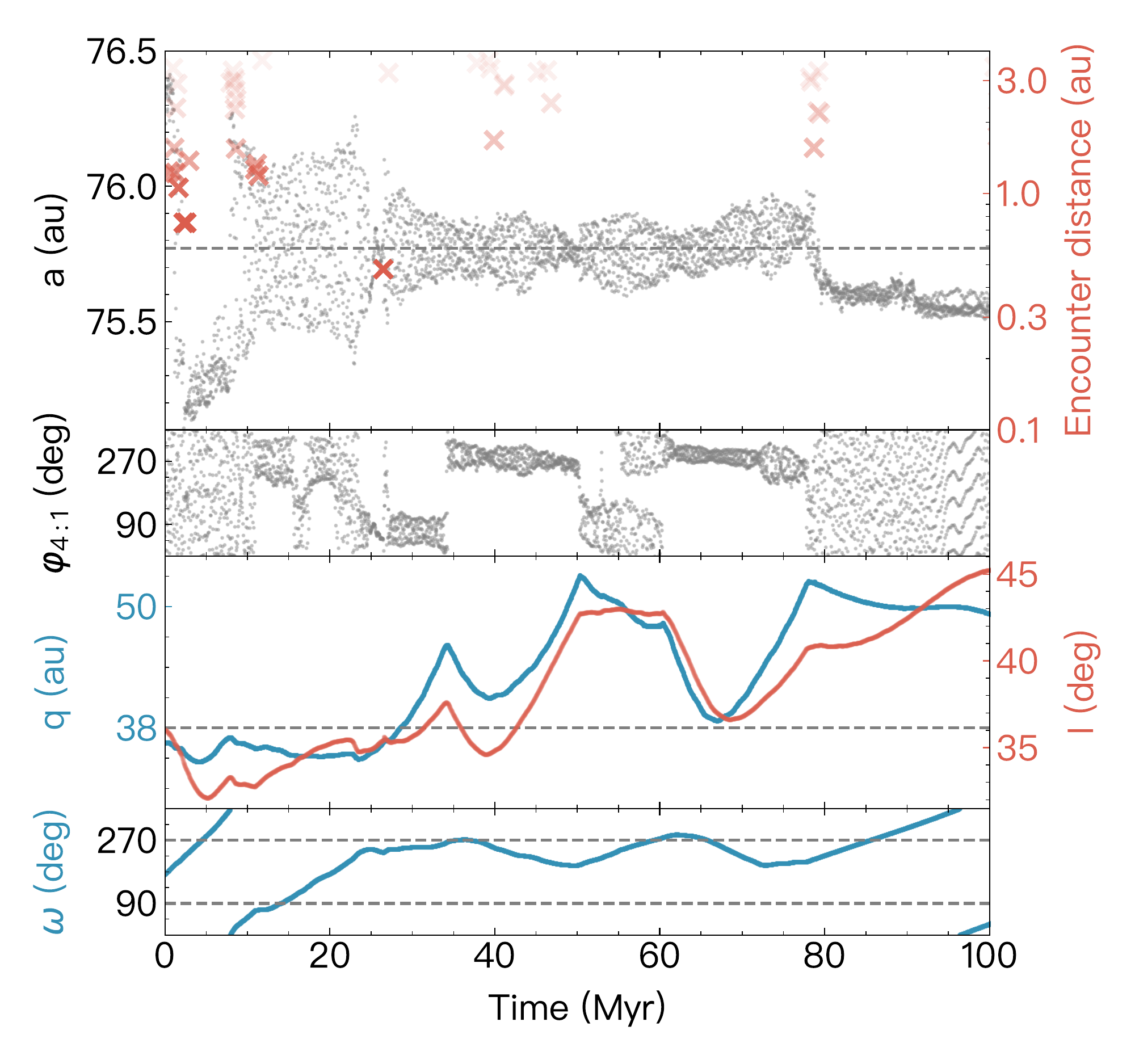}
      \label{subfig:par_rogue}
    }
    \caption{Dynamical evolutions of two particles initially near the 4:1 resonance from the reference simulation (left) and the rogue planet simulation (right). From top to bottom, the left axes of both plots are semimajor axis ($a$), the 4:1 resonant angle ($\varphi_\text{4:1}$), perihelion ($q$), and argument of perihelion ($\omega$). The right axes are proximity of rogue encounters (denoted by red `x's) and inclination ($I$). Without the rogue present, the Kozai mechanism (represented by $\omega$ librating around $90^\circ$ or $270^\circ$) is able to raise resonant particle's perihelion, but spontaneous decoupling from the resonance is highly unlikely. In comparison, the rogue helps build the detached population by both pushing the particle into the resonance and kicking it out at high $q$.  }
\label{fig:particles}
\end{figure*}

We detail a previously unreported dynamical effect that creates detached TNOs through a combination of Neptunian resonances and rogue encounters. 
We observe that weak encounters with the rogue are continuously nudging TNOs in semimajor axes, sometimes randomly pushing them into a nearby resonance from the scattering disk. 
Similarly, rogue encounters are capable of kicking objects out of the resonance; if this happens to occur at high perihelion after part of a Kozai cycle, it naturally forms detached objects near resonances, especially near those with strong $q$ lifting effectiveness like $n$:1 and $n$:2. 
Both the resonant `pushing in' and `kicking out' happen; our simulation shows the net effect is a $3\times$ enhancement to the resonant population (compared to Fig.~\ref{fig:ref}'s reference simulation), in addition to the considerable quantity of detached TNOs formed around resonances. 
The power of the rogue-aided $q$ lifting is visible in the deficits of scattering objects (gray dots and upper histogram in Fig.~\ref{fig:main}) at the resonant semimajor axes. 
Detached objects with $a < 80$ au and $q > 40$ au seem to concentrate near resonances, as do real detached TNOs (the movie illustrates these dynamics clearly). 
The resonance works as a sort of `water fountain', constantly pumping the particles to higher $q$; meanwhile the rogue supplies particles from the scattering disk, and `splashes' them to nearby detached states along the resonances. 
Such fountain-like structures with a central resonant population and surrounding detached population are visible near strong resonances like 5:2, 3:1, and 4:1.

We selected a particle near the 4:1 from each of the two simulations and plot their evolutions  (Fig.~\ref{fig:particles}). 
The reference simulation's particle (Fig.~\ref{subfig:par_norogue}) 
is {\it initially} inside the 4:1 resonance. 
It demonstrates a (rare) $\sim$25~Myr Kozai cycle, enabled by $\omega$ remaining near $90^\circ$ initially and
diagnosed by strong $e$ and $i$ coupling; this lifts $q$ from 37 au 
to 50 au and $i$ from $20^\circ$ to above $30^\circ$. 
Once Kozai stops ($\omega$ circulates), the 
still-resonant particle's critical angle jumps back and forth
chaotically between the two asymmetrical libration centers \citep{Morbidelli.1995ff}, 
with $q$ and $i$ remaining high. 
However, without additional disturbance (from a jumping Neptune or external rogue), it is almost impossible to spontaneously 
decouple from the resonance and thus become detached. 

Fig.~\ref{subfig:par_rogue} shows a case from the rogue scenario, but here the TNO is initially near but not inside the 4:1 resonance. 
Each red cross (top panel) marks a time and encounter distance with the rogue. 
These encounters 
nudge the particle's $a$, thus changing its resonant dynamical behaviour. 
At $\approx$10~Myr, weak encounters move the particle into the 4:1, beginning $\varphi$ libration around $\sim$270$^\circ$. 
Interestingly, little $q$ and $i$ evolution occurs until a deep encounter pushes the TNO into a part of the resonant parameter space where Kozai activates. 
After several $q$ and $i$ oscillations from 25--80~Myr, additional encounters at $\approx$2~au distance kick the object out of the resonance, leaving $q\approx50$ au and $i\approx45^\circ$, creating a detached TNO near the resonant border.

The juxtaposition of  Fig.~\ref{fig:particles}'s two plots 
shows three dynamical effects the rogue induces via encounters: 
(1) randomly pushing nearby non-resonant scattering disk objects into the resonance, 
(2) boosting the $q$ lifting by supplying resonant particles into the parameter space where the Kozai cycle operates, and 
(3) randomly kicking resonant particles out and forming part of the detached population if this occurs at high perihelion. 

Only a small faction of the $N$ scattering objects (that intersect rogue's orbit and are near the mutual node) will be affected per rogue orbit. 
Over the rogue's lifetime the entire scattering disk can have rogue encounters because of mutual precession of the 
rogue and TNO orbit. 
One can analytically estimate the accumulated encounter 
number $N_\text{enc}$ closer than
$\beta$ Hill spheres%
\footnote{Given the huge changes in the rogue's heliocentric distance,
%due to its large eccentricity, 
a time-varying Hill sphere $r\sqrt[3]{m/3M}$ (where $r$ is the solar distance when an encounter occurs) is used in both the numerical integrator and the analytical analysis. } 
as
\begin{equation}\label{eq:N_encounters}
    \dfrac{N_\text{enc}}{N} 
    \simeq 6\;  \beta^2 \left( \dfrac{T_r}{100\ \text{Myr}} \right) \left( \dfrac{m_r}{2 M_{\oplus}}  \right)^{\frac{2}{3}} \left( \dfrac{a_r}{300\  \text{au}} \right)^{-\frac{3}{2}}, 
\end{equation}
where $N$ is the number of 
TNOs between the rogue perihelion and aphelion,
and $T_r$ is the rogue's dynamical lifetime. 
For our simulations, the numerical integrator logged the $\beta<1$ encounters,
recording $\approx$5 encounters per particle in 100 Myr, in excellent
agreement.

Each encounter perturbs different TNO orbital elements; we focus on the 
rogue's effect on semimajor axes, as the random nudges in $a$ 
are what determine resonance entrance and exit.
We estimate 
$|\Delta a|$ 
for a typical rogue encounter at $\beta R_H$ flyby distance as:
\begin{equation}\label{eq:da_a_fin}
| \Delta a |
\simeq \dfrac{0.1\ \text{au}}{\beta} \left( \dfrac{m_r}{2 M_{\oplus}}  \right)^{\frac{2}{3}} \left( \dfrac{a}{50\  \text{au}} \right).
\end{equation}

\begin{figure*}[htb!]
  \centering
  \includegraphics[width=0.8\textwidth]{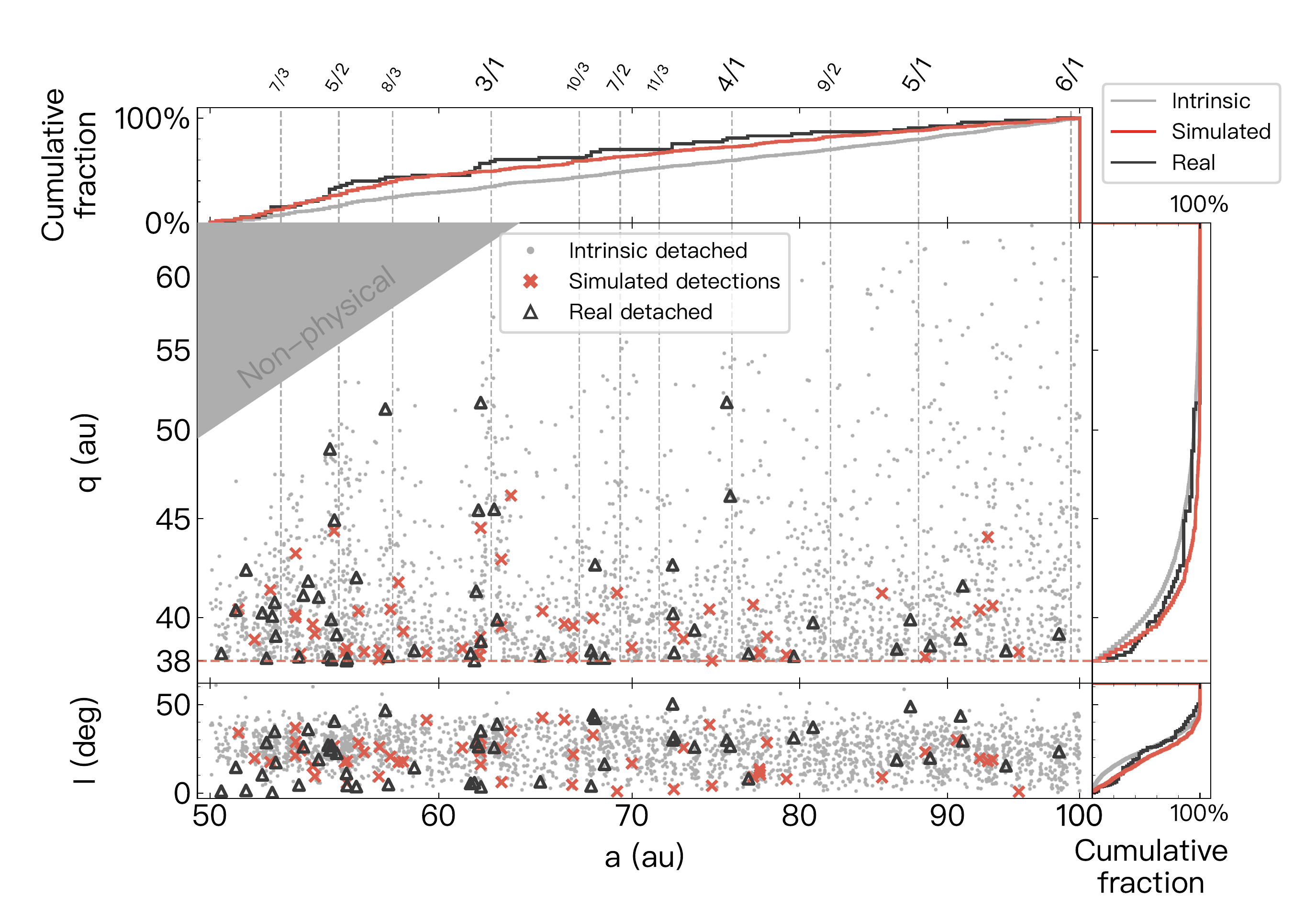}
  \caption{$a, q, i$ distributions of the detached (gray dots), the simulated detections (red crosses), and the real $q>38$~au detached objects (black triangles). The intrinsic sample is built by the rogue and has been eroded to 4~Gyr (only particles with initial $q_0 < 35$ au are included in this plot), on the basis of which the simulated detections are drawn using the OSSOS survey simulator. A decent correspondence between the simulated and the red detached can be found on the three cumulative histograms; a valid proof that the rogue is capable of creating the observed TNO distributions. }
  \label{fig:bias}
\end{figure*}

\noindent
For $a$~=~50--100~au, encounters at 1~$R_H$ induce $\Delta a\simeq0.1$--0.2~au for the TNO, approaching the $\simeq \pm$0.5~au width of the nearby resonances \citep{Lan.2019}. 
This allows encounters to knock TNOs in and out of resonance or shift them inside the resonance, allowing activation of Kozai cycling. 
Deeper encounters inducing larger $\Delta a$ do exist 
(Fig.~\ref{subfig:par_rogue}),
but Eq.~\eqref{eq:N_encounters} shows encounters become quadratically rarer 
with decreasing $\beta$.
An average TNO suffers a passage no closer than 0.4$R_H$
for Fig.~\ref{fig:main}'s rogue.

These encounters greatly increase how many TNOs end up in the high-$q$ region.
We find the rogue's 100~Myr presence raises 10\% of the $a$=50--100 au scattering disk objects to $q > 38$ au, with 3\% being resonant and 7\% being detached. 
Compared to the reference simulation, this rogue scenario emplaces an order of magnitude more TNOs in the high-$q$ region.
We also did a preliminary exploration of varying the rogue's mass; 
two additional 100~Myr simulations show that 0.5~$M_\oplus$ or 1~$M_\oplus$ rogues  still populate the high-$q$ region, but with lower  efficiency
(with 3.3\% and 5.5\%, respectively, of TNOs having $q > 38$~au). 
Both the rogue's time $T_r$ intersecting the belt and its $a_r$ and $e_r$ evolution 
history 
(Eqs.~\ref{eq:P_sec} and \ref{eq:N_encounters}) influences its sculpting 
of the distant Kuiper Belt's structure.

\section{Estimating Observation Bias}\label{sub:biases}

The real TNOs in Fig.~\ref{fig:main} are  more concentrated  to
low-$a$ and low-$q$ than the distribution produced by the rogue. 
This is expected given observational bias which favors them.
Observation biases differ from survey to survey, but the first-order effect for near-ecliptic surveys is that it penalizes large-$a$, large-$q$, and large-$i$ 
orbits. 

To verify whether our numerically simulated TNO distribution is 
similar to
the observed Kuiper Belt, we forward bias the numerical sample to compare it with the real objects. 
\citet{Lawler.2018}  details how this forward biasing is done using a `survey simulator'.
Biases for resonant objects are complex to simulate, as their perihelion passages are correlated to  Neptune's location,
resulting in detection preferentially at specific longitudes relative to Neptune \citep{Gladman.2012}. 
Lacking detailed pointing information for many past surveys, we do not compare to the resonant objects, instead focusing on $q>38$~au detached TNOs.

We first eroded the surviving test particles for 4~Gyr with only the four giant 
planets. 
That is, the rogue was assumed to be ejected after a typical dynamical lifetime of 100~Myr; in this case we manually removed it before the 4~Gyr integration. 
The dynamical classification algorithm was then repeated 
to remove resonant and scattering TNOs from the $q>38$~au sample, 
and the remaining detached TNOs 
are plotted in Fig.~\ref{fig:bias} (gray dots). 
The uniform 
initial    
$q_0$ distribution 
allowed us to weight the sample post-facto 
and we found obvious improvement in the match keeping only $q_0<35$~au (see
below).
We superpose 53 real detached objects 
(black triangles); these TNOs were identified by \citet{Gladman.2021}, consisting of the OSSOS detached \citep{Bannister.2018} and other TNOs with sufficiently 
good orbits. 
We utilized the OSSOS survey simulator 
to generate 689 simulated detections; a random 53 of them (the same as the real sample) are plotted (red crosses) to illustrate the biases.
Cumulative $a$, $q$, and $i$ histograms for $\sim$2800 intrinsic (model) particles, 
the 689 simulated detections, 
and the 53 real objects are on Fig.~\ref{fig:bias}'s side panels. 
Because detached TNOs from other surveys do not share the same detection
biases\footnote{As an example, the Dark Energy Survey's high-latitude
coverage \citep{Bernardinelli.2022} strongly favors high-$i$ TNOs.}
as OSSOS, this preliminary comparison is only approximate.

Fig.~\ref{fig:bias}'s cumulative distributions exhibit (perhaps surprisingly) good agreement between the simulated detections (red) and the real detached objects (black). 
When using all numerical initial conditions, the simulated $a$ and $q$ distributions 
have the general trend of the real detections, 
but restricting to $q_0<35$~au produces an obvious improvement.
We take this as evidence that much of the perihelion lifting began at an early stage when the scattering disk was still developing; 
Fig.~3 of \citet{Gladman.2005} shows that in the first 50~Myr only $q<35$~au
orbits are populated, only after $\sim$1~Gyr do
scattering TNOs extend up to $q=37$.
The superiority of a more confined $q_0$ distribution 
is verified in Beaudoin et al. (2022, submitted to PSJ), 
who more rigorously compares with only the OSSOS objects;
they show that the $q_0 < 35$~au detached TNO $q$ distribution created by 
the 2-$M_\oplus$ rogue is non-rejectable, with Anderson-Darling probability 
of $32\%$ 
(and is in fact the best model they studied).

\section{Discussion}\label{sec:dis}

We demonstrate for the first time that a rogue planet present for
$\sim$100~Myr during planet formation can abundantly create both the distant resonant and detached populations.
This is accomplished by the synergy of the Neptunian resonances (with the Kozai mechanism lifting perihelia) and weak rogue encounters (where the rogue supplies the resonance and detaches objects at high $q$). 
Several points merit discussion.

\begin{figure}[htb!]
  \centering
  \includegraphics[width=1\columnwidth]{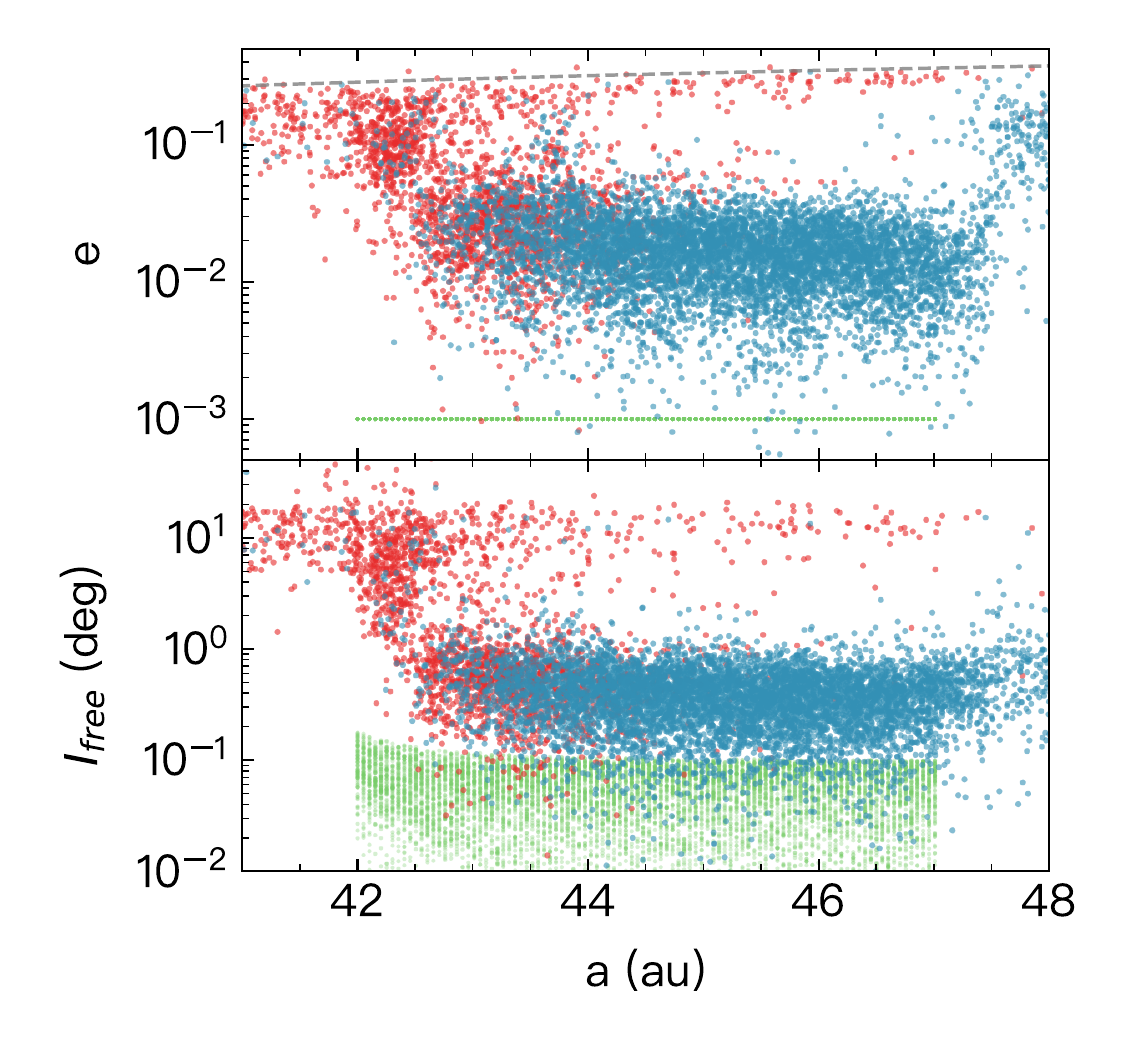}
  \caption{Orbital excitation of the cold classical Kuiper Belt, after 100~Myr of perturbation from the rogue with $a,q,i$ of roughly 300 au, 40 au, and 20$^\circ$.
  Green points mark initial conditions, and red/blue points show final values for
  particles that began with $a_0$ smaller/larger than 43~au, respectively.
  From initial $e_0=10^{-3}$ and $i_\text{free}<0.2^\circ$, 
  the surviving cold main belt  with $a_0>43$~au 
  remains with low $e$ and $i$.
  For $a_0 < 43$~au secular resonances ($g_8$ and $s_8$) pump
  both $e$ and $i$ after which particles scatter to all values of $a$ along the Neptune crossing line (gray dashed). Repeating the simulation using \textsc{Rebound} \citep{Rein.2012} produces the same excitations.
  Thus, the rogue does not unacceptably excite the cold belt, whose TNOs currently have even higher values of $e$ and $i$.}
  \label{fig:cold_belt}
\end{figure}

{\bf The Cold Classical Belt.}
A potential concern regarding the temporary presence of Earth-scale planets is the possibility of dynamically heating 
the cold classical belt, which is often thought to be 
formed in-situ and unexcited for the age of the solar system. 
The observed limits on the  $e$ and $i$  excitation  are used to constrain 
Neptune's dynamical history \citep{Batygin.2011, Dawson.2012, Nesvorny.2016ij8},
including the absence of planets formed in the cold belt itself \citep{10.1006/icar.2002.6832}.
A rogue must have scattered to large $a$ early, as 
even a few million-year
residence with $a_r\simeq$~50~au would excite the cold belt. 
Once the rogue reaches $a$ of a few hundred au, the average time it stays in the classical belt drops by $a^{3/2}$ (Eq.~\ref{eq:N_encounters}), greatly reducing the cold-belt's excitation. 
We confirmed this with simple numerical simulation of just 
the cold classical belt, placing
10,000 objects with initial $e_0=
10^{-3}$ and $i_\text{free} < 0.2^\circ$ \citep{Huang.202269} from $a_0 = 42$ au to 47 au and integrated with the same rogue in Sec.~\ref{sub:rogue}. 
Even though the rogue continuously crosses this cold 
belt\footnote{For this specific rogue, 50\% of the 100~Myr has one mutual node inside the cold belt.} 
for 100 Myr,
its gravity induces surprisingly little excitation:
the vast majority of cold TNOs keep $e < 0.05$ and $i_\text{free} < 1^\circ$ (Fig.~\ref{fig:cold_belt}). 
We conclude a large-$a$ rogue
does not unacceptably excite the cold belt.

{\bf Oort Cloud Building.}
The period of the rogue's existence will coincide with the epoch in which the
Oort cloud is created
\citep{Duncan.1987, Dones.2004, Zwart.2021}. 
Although in principle one might worry that an Earth-scale rogue at
100~au could strongly interfere with the creation of the Oort cloud,
\citet{Lawler.2017} show that the presence of even a larger 10~$M_{\oplus}$ 
object in the 250--750~au range for the entire age of the Solar System
lowers Oort cloud implantation efficiency by only a factor of $\simeq$2.
The efficiency of Oort cloud implantation and its mass are sufficiently uncertain \citep{Zwart.2021}
that there is no obvious problem with the temporary presence of a rogue 
like we envision.

{\bf Sun's Birth Environment.}
The rogue's highly-eccentric orbit of a few hundred au could be affected by close-in stellar flybys that could have happened in the Sun's birth environment. 
Arguments have been made that our Sun was likely born in a cluster of 1000-3000 stars/pc$^3$, based on extinct radionuclides and the assumption that extreme detached TNOs, represented by Sedna, needed to produced in the birth cluster environment 
\citep{Zwart.2009, Adams.2010, Pfalzner.2013}. 
The question is how quickly the Sun exited this birth cluster.
If Sun remained for a long time, there would be problems with retaining the Oort cloud \citep[\textit{egs.}][]{Morbidelli.2004, Nordlander.2017}. 
In addition, the recent study by \citet{Batygin.2020jr} computes an upper bound of number density-weighted cluster residence of our Sun of $2\times10^4$ Myr/pc$^3$, based on the unexcited inclination distribution of the cold classical belt; 
this implies that the Sun must have exited its birth cluster in less than
$\sim$15 Myr,  
using their 1400 stars/pc$^3$ estimate \citep{Batygin.2021}. 
Similar early exit arguments are given by \citet{Brasser.2006} and \citet{Pfalzner.2013}, both of which suggest 5 Myr residence. 
The timescale for the rogue to reach several hundred au is comparable to this $\sim$10 Myr duration and we therefore think a 100 Myr survival timescale for the rogue is not problematic. 
Furthermore, the rogue's presence directly provides a way other than the Sun's birth cluster to explain Sednoids 
\citep{Gladman.2006}, 
which would alleviate the ``need to make Sedna with passing stars" constraint 
\citep[fig.~7 of][]{Adams.2010, Pfalzner.2013, Brasser.2015} 
in the Sun's birth environment.

{\bf An Existing Planet.}
The natural 100~Myr ejection time scale for scattering rogues \citep{Gladman.2006} sets a typical time scale that we have seen
produces the needed detached and resonant populations in the 
50--100~au region.
Instead of ejection, if the rogue's perihelion was lifted (by
an unspecified process) to very large $q$, it 
could  remain in the outer solar system today and
negligibly affect the  50-100~au region.
Scenarios with a still-resident rogue \citep{Sheppard.2016, Batygin.2019}
presumably began with that planet on low-$q$ orbit  
for some period; that combination of $T_r$, $m_r$, and $a_r$ (Eq. \ref{eq:N_encounters}) while $q_r<100$~au could produce the same
effects we study, before the mysterious $q_r$ lift.
But given that recent surveys \citep{Shankman.2017cuc, Napier.2021, Bernardinelli.2022} do not support intrinsic clustering, we find the `now gone' rogue scenario
to be more natural.

{\bf Neptune Migration.}
 It is generally believed that Neptune migrated outwards during the planet-formation and disk-dispersal epoch \citep[reviewed by ][]{Nesvorny.2018}.
 The `grainy' migration models (see Sec.~2) are effective at creating detached TNOs when Neptune's semimajor axis jumps (and thus so do all its resonances) due to encounters with dwarf planets; if the jumps become comparable to the resonance size ($>$0.1~au, say), then  some particles are suddenly no longer in resonance. 
 A final phase of slow net-outward grainy migration results in an asymmetry of `stranded' particles on the sunward side of the resonance
 \citep{Kaib.2016l9v, Nesvorny.2016}.  
 There is growing observational evidence for this
 \citep{ Lawler.2019, Bernardinelli.2022};
 after fixing an error in the Dark Energy Survey selection function (working with Bernardinelli, private communication 2022), we combined the $q>38$ au
 samples from these two studies and find that the binomial probability that the detached number just beyond each resonance is comparable to those on the sunward side remains $<$1\%.
 
 Regarding detachment, we find the rogue planet scenario produces comparable numbers of detached TNOs.  
 This is not too surprising, if one takes the view that the rogue produces `grainy' TNO jumps while grainy migration jumps the resonances.
 In our case, 
 Eq.~\eqref{eq:da_a_fin}'s $\Delta a$ is set by the range of encounter distances and the rogue's mass, while in grainy migration models there is an assumed mass spectrum of the bodies encountering Neptune (at a range of flyby distances).
 It is likely that after the rogue's ejection there will still be 
 moderately-massive scattering disk; during its erosion
 a final small outward Neptune migration will then occur.
 This would capture stranded TNOs on the high-$a$ side of the
 resonance and continue `littering' TNOs on the sunward side, giving an outcome very similar to migration alone.
 
 A unique outcome of our study is that we have rigorously compared the simulation's final orbital distribution to the known OSSOS TNOs,
 and find excellent agreement (Beaudoin et al. 2022, submitted to PSJ),
 yielding a population estimate of 40,000 detached TNOs with diameters
 $>$100~km, a number identical to the \citet{Nesvorny.2016} population estimate, who didn't have the information necessary for rigorous orbital comparison.
 Additionally, having a rogue during this period simultaneously allows the production of large-$q$ objects like Sedna \citep{Gladman.2006}.
 
 We believe that both processes operated in our early Solar System
 because it is natural that objects between Pluto and ice-giant scale existed during disk dispersal.
 The rogue's presence introduces another mechanism to produce many features seen in the distant Kuiper Belt.
 We believe that rogues and migration are both expected outcomes of the process of planet building; the uncertainties introduced into deriving parameters (such as the migration duration and mass spectrum of other bodies in the system) in future
 models which incorporate both seem unavoidable.

\section{Acknowledgements}

We thank F.~Adams, P.~Bernardinelli, L.~Dones, , S.~Lawler, S.~Tremaine, K.~Volk, and an anonymous referee for useful discussions. YH acknowledges support from China Scholarship Council (grant 201906210046) and the Edwin S.H. Leong International leadership fund, BG acknowledges Canadian funding support from NSERC.

\bibliography{_ref}{}
\bibliographystyle{aasjournal}

\end{document}